\documentclass[10pt]{article}
\usepackage{graphicx}

\usepackage{epsfig}
\usepackage{latexsym}
\usepackage{amsmath}

\begin{document}

\title{ Another analytic view about quantifying social forces. 
}

\author{Marcel Ausloos\footnote{
$e$-$mail$ $address$:
marcel.ausloos@ulg.ac.be; {\it  previously at } GRAPES@SUPRATECS, Universit\'e
de Li\`ege,  Sart-Tilman, B-4000 Li\`ege, Euroland
} \\   R\' esidence Beauvallon,
 rue de la Belle Jardini\`ere,  483/0021 \\B-4031, Angleur, Belgium }

 \date{\today}
\maketitle
\begin{abstract}
Montroll   had considered a Verhulst  evolution approach for introducing a notion he called "social force",  to describe  a jump in some economic output when a new technology or product outcompetes a previous one.   In fact,  Montroll's adaptation of Verhulst equation is more like an economic field description than a "social force". The empirical  Verhulst logistic function and  the Gompertz double exponential law are used here in order to present  an alternative  view, within a similar mechanistic  physics framework. As an example, a "social force" modifying the rate in the number of temples constructed by a religious movement, the Antoinist community,  between 1910 and 1940 in Belgium is  found and quantified. Practically, two temple inauguration regimes  are seen  to exist over different time spans, separated by a gap attributed to a specific "constraint", a taxation system, but allowing for a different, smooth, evolution rather than a jump. The impulse force duration is also emphasized  as being better taken into account within the Gompertz framework. Moreover, a "social force"  can be  as here,  attributed to a change in the  limited need/capacity of some population,  coupled to some external field, in either Verhulst or Gompertz equation, rather than   resulting from already existing but competing goods as imagined by Montroll.
\end{abstract}



\maketitle

\section{Introduction}\label{intro}

Malthus  \cite{Malthus798}  had assumed that the growth of a biological population was strictly proportional to the number of members in the population, $dN/dt  =  r\;N$, thereby obtaining an unreasonably excessive exponential growth $N(t)\simeq e^{rt}$.    To correct  for such a behavior, Verhulst attempted to find an evolution equation  with  a limiting growth term \cite{Verhulst845,Verhulst847}.   He  empirically introduced a quadratic term, in the right hand side of Malthus equation, in order to mimic the food availability and/or the so called  carrying capacity $M$ of the country, 
 \begin{equation}\label{modN}
\frac{dN}{dt} = r N \left( 1- \frac{N}{M} \right).
\end{equation}
He obtained the famous, sigmoid, logistic function.
  In contrast,  Gompertz    \cite{Gompertz825}   formulated   the decrease in the number of members of a population   in terms of a  first order differential ($growth$) equation of Malthusian form,  but with an {\it exponentially decaying birth rate}, $ r= r_0 e^{-\kappa t}$.  This  results into another analytical sigmoid curve,  though asymmetric,  in contrast to the symmetric  logistic map. 
  
 In 1974, Montroll and Badger \cite{MontrollBadger74}  had  introduced, into social phenomena,  quantitative approaches,  as physicists should do along mechanics lines. 
Nowadays, one would say that (i)   "degrees of freedom" may  interact  with ''external fields'', like a spin   interacts with a magnetic field or a charge interacts with an electric field, $and$ (ii)  "degrees of freedom" may  interact with each other within some cluster, e.g. like for a spin-spin interaction.
 
 It was  observed, e.g., in   \cite{PNAS75.78.4633-7-Montroll-socialforces},  that deviations from the logistic law  \cite{Verhulst845,Verhulst847} were often associated with unusual
intermittent events: wars, strikes, economic panics, etc. Mathematically, the effects
can be abstracted
through an instantaneous   impulse function, called a "social force", by Montroll  \cite{PNAS75.78.4633-7-Montroll-socialforces},  inserted  into Verhulst equation.     In so doing,   some remarkable shift  occurs in the  time evolution  of events,  after  some  delay, - the shift size depending on the force/impulse strength.   Montroll presented semi-log   graphs,  with such apparently drastic shifts, on 
 the replacement of rail service by air service in intercity passenger travel, or the replacement of sailing ships by steamships,  - both rather "economic conditions", or  an accelerating  decline in agricultural work force, - an effect apparently due to telegrams from the White House. 
 
 However, the notion/definition of some "social force", as in 
Montroll  \cite{PNAS75.78.4633-7-Montroll-socialforces},  is  rather loose. Nowadays, the notion of "social field" seems to be a more realistic concept \cite{Helbing1,Helbingbook}. In fact, Montroll discussed  industrial evolutionary processes,  considered as a sequence of replacement or substitution  processes.  It is like a competition between two  equilibrium states in a free energy-like description. In such a spirit, shifting the "thermodynamic state" toward one or another minimum,  is rather due to the rise of an "economic (potential)  barrier";  thus a wording like  "economic force" could have rather been used.   

Incidentally, the analytical form  of the force impulse as introduced by Montroll into Verhulst equation, see Sect.  3, appears rather {\it ad hoc}. Moreover, it only allows for a constant and identical growth rate on both sides of some time interval, - as that was the case in the data presented  by Montroll.

In view of  such considerations, I was interested in finding whether one could get some   "social data" on the evolution of some  community, made of agents having a well defined   ''degree of freedom'',  asserted to some external field, and whether one could observe some (keeping the same wording as Montroll) "social force" through appropriate data analysis. Moreover, to find smoother evolutions  and various growth rates should be a plus.  This  would request  more  high frequency  data than  yearly ones, as  in [1]. 

 An indirect measure of social behavior, like the   construction and inauguration/consecration of temples by a religious movement,   seems  reliable, more than, e.g.,   adhesion numbers as in  \cite{religion1,mendes,religion2,religion568,hayw99,hayw05}, - relying on statistics based on interviews and surveys.  Moreover the former data involves more immediate mixing of economy and social conditions than the number of adepts.  
  
As a practical example, the  evolution of the number of temples, built in Belgium,  
by a community like the Antoinists \cite{MAarXiv1201.4841}
is  reported in Sect. \ref{sec:dataset} and    
 first studied   through  the most commonly accepted  kinetic  growth law, i.e.  the Verhulst logistic function  \cite{Verhulst845}; see Sect. \ref{ANTEv}.

Waiting for a change in taxation laws in Belgium,  the construction of temples stopped during a while, after the end of the first World War. When the constructions resumed, the growth rate apparently changed.  Montroll's "formalism" is thus $ inappropriate $, on two counts at least. In fact, it will be seen, in Sect. 3,  that one can introduce another  analytic form of a  "social force" into an evolution equation in order to describe a rate change, - with or without a gap in the evolution. 

Moreover,  the Verhulst mapping is sometimes criticized as unrealistic. The    Gompertz (human mortality) $decay$ law   \cite{Gompertz825} is often suggested to be the alternative, -   tuning the parameters into their size $growth$ value, as often done nowadays.  Some quantitative analysis is thereby found in Sect. \ref{ANTEg} along such a line of approach.  
Thus, quantitative measures, as expected within Montroll and Badger pioneering  framework  \cite{MontrollBadger74}, 
are obtained. They are commented upon in  Sect.\ref{summary}.

 In fact, Verhulst  and Gompertz basic evolution equations are particular examples of a more general first order differential equation  \cite{MaBio179.02.21} which is discussed in the conclusions, Sect.\ref{Conclusions}, with the idea of presenting the general addition of  an external field-like term in such a framework.
 
   In brief,  this study has a marked difference from   that of Montroll. It is not a competition effect  between two processes, that one is  looking at here below, but rather an evolution due to an external field, - a taxation system here.  Here, the "social force" is  attributed to a change in the  limited need/capacity, in Verhulst or Gompertz equation, rather than   resulting  from already acquired goods competing with new ones.Thus, it can be pretended that  one can be  presenting two different views of a "social force", - as resulting from two different origins of the minima in some sort of thermodynamic free energy potential well description:  a competitive interaction or the effect of an external field, leading to the choice of the (kinetic but equilibrium-like) state. 
   
   Moreover, in the conclusion section,  the mathematical form of both types of forces   receives some sociological interpretation. This can be based on the change in the  limited need/capacity of some population,  coupled to (or influenced by) some external field,   rather than   resulting from already acquired but competing goods as in [1].

 \begin{figure}
\centering
\caption{Logistic fits,  at low and high $t$,  of the number of temples of the Antoinist Cult  in Belgium as a function of the number of months  since the consecration of the first temple on Aug. 15, 1913 in Jemeppe-sur-Meuse.  The "size" of a year is of course 12  months. Each year is "centered" on July 01.  N.B. "1910" on the figure is a misprint: it should read 1913.}
\includegraphics[height=20cm,width=16cm]
{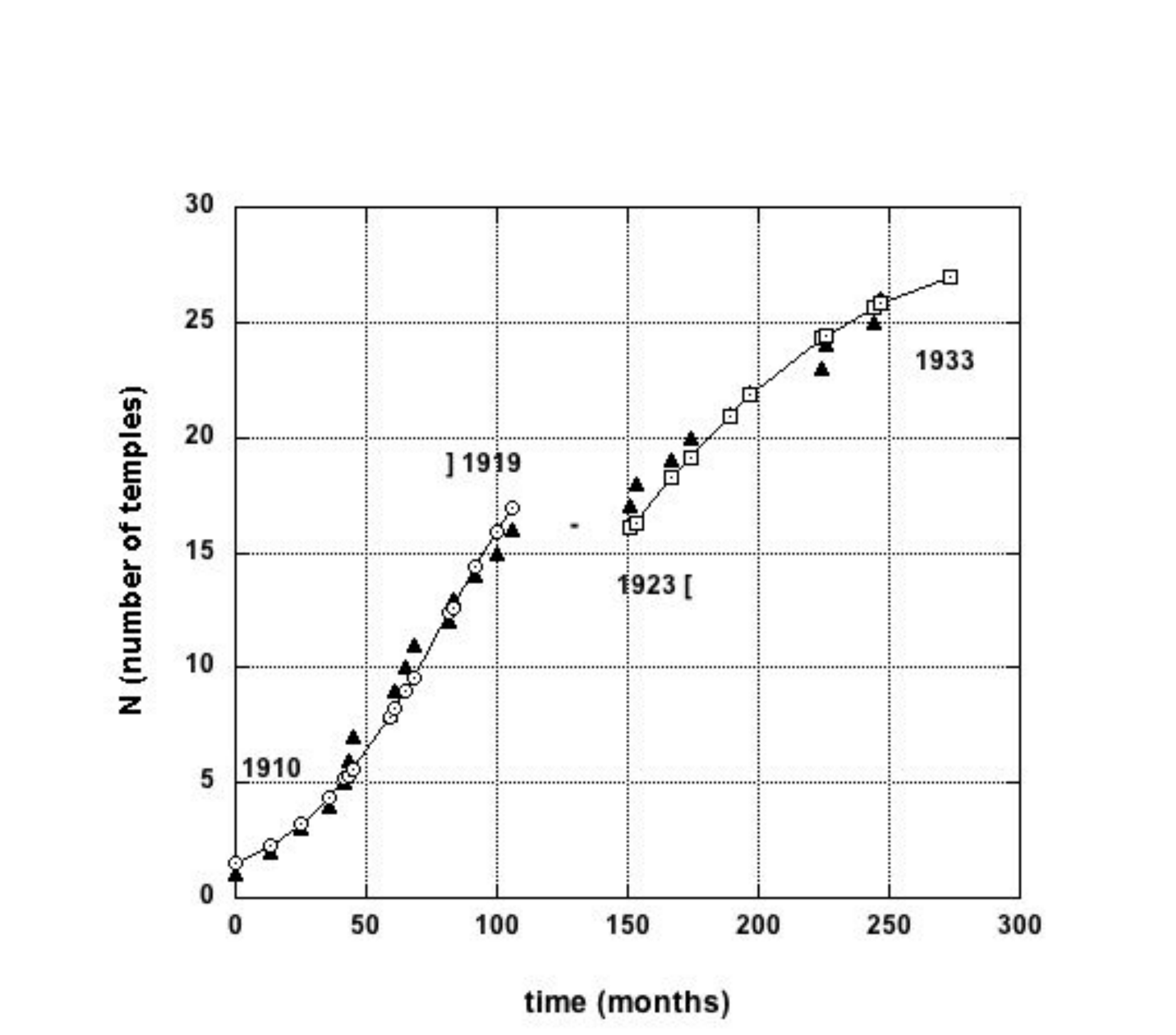}
\label{fig:Plot30ALMFtempleVfits}
\end{figure}

\section{  Data Set}
\label{sec:dataset} 

  The Antoinist community exists for about more than a century   \cite{Debouxhtaybook,Dericquebourg}, has markedly grown, has expanded in various countries, but and is now apparently decaying. Although it might  be of interest to consider the data for the whole "sect" on a world wide basis, only the 27 temples constructed in Belgium during the  original growth phase of the  cult  are here  considered.  The date of  every temple consecration has been extracted from the archives of the Antoinist cult and from \cite{Debouxhtaybook} and \cite{Dericquebourg} compilation and discussions.    Most of the times, the exact day of the   consecration  is known, -   twice, only   the month is known. In such a case, I took, the 15-th day of the month as the consecration day. To be more precise about  the exact day of  the event, for both cases, has requested much time consuming research for this information through news media archives,  - but without  any  success. Sometimes, some disagreement,  between dates, was found. When in doubt, the dates in an Appendix  of a 1934 book by Debouxhtay \cite{Debouxhtaybook} were preferred, since I consider them as the most reliable ones,  - due to some ''proximity effect'' of the book's author .
   
 In order to keep the same units, after calculating the number of days between two events, the number  ($m$) of months  (months) has   been next rounded up to the nearest integer.  
 
Thereafter, the evolution in the number of temples  was analyzed in terms of the Verhulst (logistic) law and the Gompertz (double exponential) growth law. In both cases, it  readily appeared  that two regimes  had  be investigated: (I) one between 1910 and 1919,  for 16 temples; (II) another between 1923 and 1935, for 11 temples.
 For further reference note that   the  $\chi^2_{15}$ value 
with 15 degrees of freedom and the $\chi^2_{10}$ with 10 degrees of freedom
have a critical value equal to 24.996 and 18.307  respectively, for a 0.05-level test  \cite{chi2table}.

\begin{figure}
\centering
\caption{
 Gompertz double exponential law  fit of the number of consecrated Antoinist temples in Belgium as a function of the number ($m$) of months (cumulated), in  low and high $t$ regimes  since the consecration of the first temple on Aug. 15, 1913. The "size" of a year is of course 12  months. Each year is "centered" on July 01. N.B. "1910" on the figure is a misprint: it should read 1913.
 }
 \includegraphics[height=20cm,width=16cm]{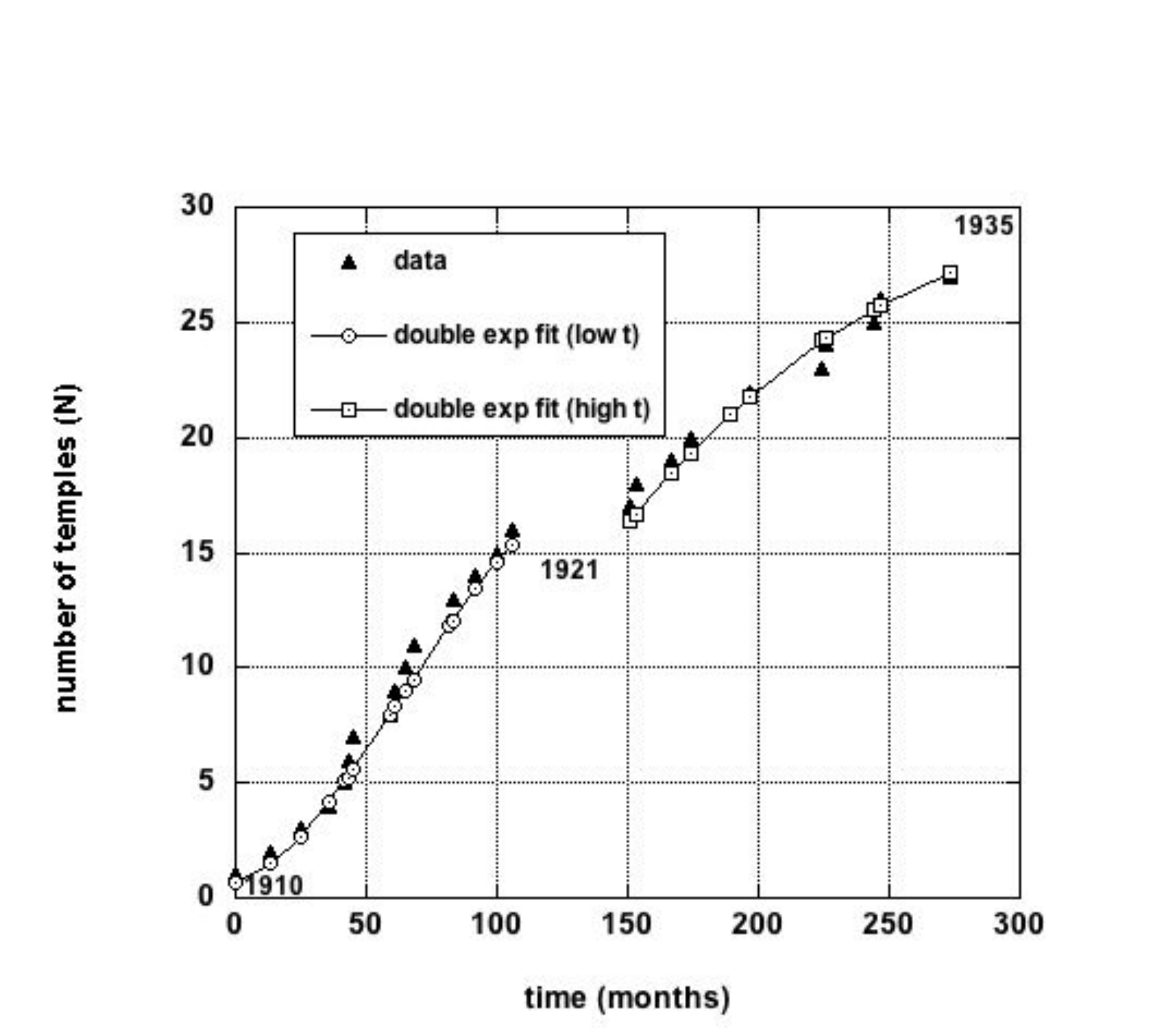}
\label{fig:Plot31AKLFtempleGfits}
\end{figure}

\subsection{Evolution study. Verhulst equation  }\label{ANTEv}

 
 
 Let us take the (Verhulst) logistic  function, with $N/M\equiv z$,  as a first approximation of a growing entity, i.e.
the so called  logistic map, a sigmoid curve, 
\begin{equation} \label {Verhulst2sol}
z(t)= z_{\infty} \frac{e^{r(t-t_0)}}{1+e^{r(t-t_0)}}\; =  \frac{z_{\infty}}{1+e^{-r(t-t_0)}}\;  ,
\end{equation}
 where $z_{\infty}$  is the upper limit  of $z$  as time tends to infinity, $t_0$ is  the position  of the inflexion point   and $r$ is the supposedly constant growth rate.  This way of expressing the logistic curve has the advantage that the initial measure, here $z_0=1$ at $t = 0$, is a rapidly fixed value for one of the  three model parameters.

 The upper value $z_{\infty}$  is  {\it a priori} imposed to be an integer.  
The low $t$  logistic fit of the number  ($N$) of temples as a function of the number ($m$) of months (cumulated since the rise of the first temple, on Aug. 15, 1910) 
corresponds to
\begin{equation}
\label{Vlowt}
   N(m) = 24/(1+e^{-0.03395*(m-80)})
\end{equation}
 while the fit in the upper  regime corresponds to
 \begin{equation}
\label{Vhight}
  N(m) = 29/(1+e^{-0.0195*(m-140)}).
\end{equation}
Both data and fits, in the appropriate regimes, are combined and shown  in  Fig. 1. For statistical completeness, let the $\chi^2$ values for the fits be reported: they are respectively  1.208 and 0.392, much below the critical 0.05 test value, recalled here above.

It seems worth pointing out  that the initial growth rate   is about  0.03395, i.e. largely more than 3 temples every ten years, but  is reduced to 0.0195 in the latest years, i.e. about 2 temples per year.  Note that a  $unique$ logistic curve would give a value of the  growth rate $\sim$ 0.02355.  The curve is not displayed since such a value does not fulfill the Jarque-Bera test.   

 




\subsection{Evolution study. Gompertz equation  }\label{ANTEg}

 Gompertz's  law is also  a 3 parameter expression, i.e.
\begin{equation} \label {Gompertz2sol}
y(t) / y_{M} = \; e^{ -b \;exp (-rt)},
\end{equation}
where $y_M$,  $b$ and $r$ are positive constants.   This corresponds to  an exponentially $decaying$ birth rate $r$ in Malthus equation, i.e.
$ r= r_0 e^{-\kappa t}$, pending $r_0$ and $\kappa$ being positive constants. 

The  derived differential equation 
 is commonly referred to as the Gompertz equation,  i.e., 
\begin{equation} \label {Gompertz1dif}
\frac{dy}{dt}= r \; y \;  log\left[  \frac{k}{y} \right]\;\\
\end{equation}
where $k$ has {\it mutatis mutandis}  the same meaning as $M$, the carrying capacity, in the Verhulst approach. 

 Fits to  Gompertz double exponential law  can be searched for  with different techniques  \cite{2280354fits,2280430fits Farebrother,Norton76NatureGompertzgrow,2347021comparingVG,3315118fits,2583955fits,PhA368.06.225Gompertzrosu}, still imposing the amplitude  $y_M$ to be an integer. It has occurred  after much simulations that  two distinct regimes must be considered, exactly as in the analysis along Verhulst  approach: one at {\it low} time, i.e. during the initial growth of the Antoinist communities, and another at later ($high$)  time, with a 4 year gap, between 1919 and 1923: 
\begin{equation}
\label{Glowt}
   N(m)=23\;e^{-e^{-(m-62)/48.5}}
\end{equation}
 \begin{equation}
\label{Ghight}
  N(m)=31\;e^{-e^{-(m-116)/77.5}},
\end{equation}
  respectively, as shown on Fig. 2. 
   Note that the upper  ($absolute$) values of the possible number of temples which could  be asymptotically expected, slightly differ in  either Verhulst or Gompertz approach. 

Recall that   the logistic law leads to expect $\sim$ 24 temples at saturation,  for the initial regime, but  29 temples at most, for the second regime. These values are 23 and 31 respectively for the Gompertz law fits.  
     
For statistical completeness, let the $\chi^2$ values for the fits be reported: they are respectively 1.480 and 0.249, much below the critical 0.05 test value again.

 \begin{figure}
\centering
\includegraphics[height=14cm,width=16cm]{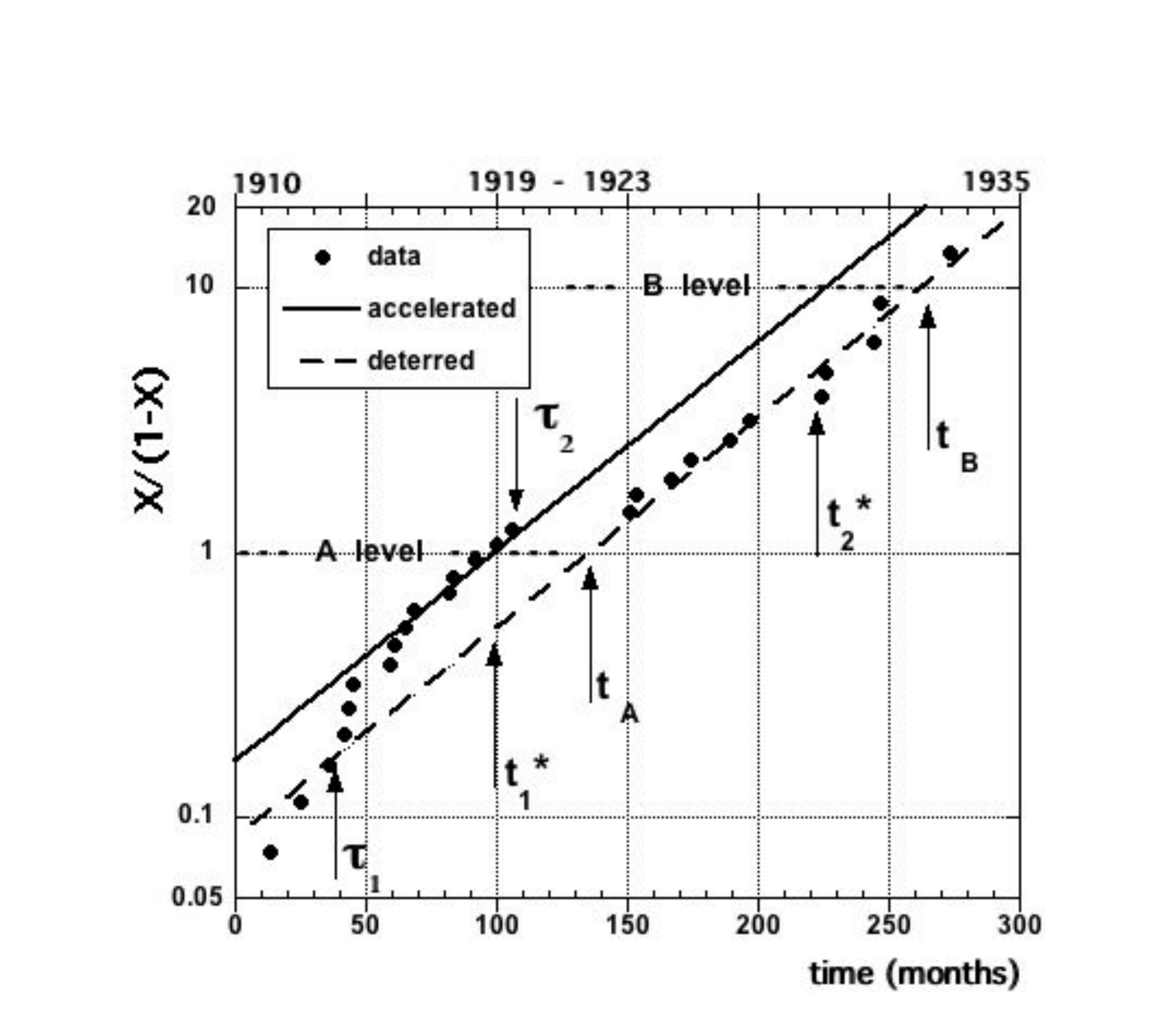}
\caption{Logistic variation    ($X/(1-X)$) of the number of temples ($X$)  in Belgium as a function of the number of months (cumulated from the raise of the first temple), in order to indicate the effect of a "social force" at time $\tau$, accelerating the process over a time span $\Delta t_A  \equiv (t_A-t^*)$ allowing to measure the strength $\alpha$ of the force impulse   in the sense of Montroll.  }
\label{fig:Plot53templeVforce}
\end{figure}

\section{ Quantitative measure of  "social forces"}\label{summary}

At first sight, the  growth regimes do not seem  to overlap much, to say the least. They occur on different sides of a time gap. Moreover, the rates of growth seem somewhat different in the successive regimes, indicating  sequential rather than overlapping (or/and competitive)  processes,  as in Montroll report. 

 It is worth to recall that Montroll argues that social  evolutionary processes occur as  a sequence of  new ideas for old ones, inducing deviations  from the classical logistic map associated with  intermittent events. 
 Montroll has argued that the most simple generalisation of Verhulst equation,  in such a respect, goes when introducing a force impulse,  $F(t) = \alpha\;\delta(t-\tau) $, in the r.h.s.  of  the kinetic  Verhulst equation   appropriately  rewritten   to emphasize the growth of the logarithm of $X \equiv   z/z_{\infty}$,
\begin{equation}
\label{Veq}
\frac {d X}{dt }= kX(1- X) , \; \rightarrow  \;  \frac {d\;ln (X)}{dt }= k(1- X) ,
\end{equation}
 so that the dynamical equation for some instantaneously forced, at time $\tau$,  evolution process $X$  reads \cite{footnote}
\begin{equation}\label{Montrolleq}
\frac {d\;ln(X)}{dt }= k(1- X) +   \alpha\;\delta(t-\tau) .
\end{equation}
In other words, 
\begin{equation}\label{MontrolleqV}
\frac {dX}{dt }= kX(1- X) +   \alpha\;X\;\delta(t-\tau) .
\end{equation}

In so doing, in the time regime after the withdrawal of  the intermittent
force, the   evolutionary curves are $parallel$ lines, on a semi-log plot:  the unaccelerated one being above or below the latter depending whether the process is accelerated or deterred at time $\tau$.    The impulse parameter $\alpha$  is obtained from the  step-like shift   \cite{PNAS75.78.4633-7-Montroll-socialforces} 
 to be
\begin{equation}
\label{alpha}
\alpha = k \; [1- X(\tau)]\; \Delta t_A 
\end{equation}
where $\Delta t_A  \equiv (t_A-t^*)$ measures the time which has been gained (if $\Delta t_A  \ge 0$)  in reaching some (arbitrary) $A$ level, i.e. $X_A/(1-X_A)$.

Two such  impulse effects can be seen in Fig. 3, which is an appropriate replot of Fig.1. An acceleration at $\tau_1   \simeq  36$ (months) and a deceleration after  $\tau_2 \simeq  106$ (months).  One easily calculates  that  $X(\tau_1)=0.14$, since  $X(\tau_1)/(1-X(\tau_1))= 0.16$ and   $X(\tau_2)=0.55$, since  $X(\tau_2)/(1-X(\tau_2)= 1.23$.  To estimate $k$, note that 125 months  were
required for $X/(1-X)$  to be multiplied by a factor 10 in
evolving from 0.1 to 1.0  or from 1.0 to 10.0, so that $k= (1/125)\; ln (10) = $\;0.0184/month. Moreover, $ \Delta t_A = -\;\Delta t_B = 40$ months. Hence,   the  impulse strengths are:  $\alpha_1= 0.103$ and $\alpha_2=  - 0.405$. These are very reasonable orders of magnitude  \cite{PNAS75.78.4633-7-Montroll-socialforces}.

 It should have been obvious  from Fig. 1 that the decelerating force strength should be higher in  absolute value  than the 	accelerating one.  

However, on one hand, there is no known reason implying that a "social force" be instantaneous and have no duration. Moreover there is no reason that the two growth rates, before and after an impulse, be identical, - the more so if the pulse has a finite duration $\theta$.

A different  adaptation of the ideas in [1],  on the evolution of competing entities, economic or sociologic ones,  occurs if one writes
\begin{equation}
\label{Montrolleq2}
\frac {d\;ln(X)}{dt }= k(1- X) +  ( \alpha /\;\theta) (1- X) 
\end{equation}
instead of  Eq.(\ref{Montrolleq}). In other words, one is (mathematically)  letting Montroll's $\alpha$ to be $X$-dependent over the time interval [$\tau; \tau+\theta$].  However, the emphasis  differs much from [1]: rather than modifying the (Malhtus) $X$ term, one adapts the (Verhulst) $(1-X)$ term to (economic or social) constraints.

Mathematically, on the same footing as  Eq.(\ref{Montrolleq2}), one can write 
\begin{equation}
\label{Montrolleq2V}
\frac {dX}{dt }= kX (1- X) +  ( \alpha /\;\theta) X (1- X).
\end{equation}

Readily, the rate before the pulse  is $k$ but is $k+\alpha/\theta$   after the "pulse" application.  This "second rate" depends on the pulse strength and some  time duration.

Furthermore, the idea can be easily carried over  within the Gompertz framework. It is "sufficient" to replace $(1-X)$ by $\sim -ln(k/y)$; see Eq.(\ref {Gompertz1dif}).
A double log plot for $N/N_M$ is shown in an appropriate replot of Fig.2, i.e. Fig. 4, - the best fit  equations being  written in the figure. The fit is very precise. The lines are $not$ parallel anymore. From these, one can deduce $\theta_1= 34$ (months)  and $\theta_2= 135$ (months), i.e. $\sim$ 3 and 11 years respectively.

 One can briefly elaborate on such observations. In fact,  the  cult  present  hierarchy interpretation  and mine go along   historical lines.   The first acceleration, ca. 1914  can be historically connected to the first World War. 

After the war,  income and housing taxes  were implemented in Belgium, but social organizations  were partially screened from such taxes.  It took more than $four$ years for a law on  {\it Organizations of Public Utility} to be voted upon,   - in 1922. The parliamentary delays  have {\it in fine} been much decelerating the temple construction and consecration process,  such as introducing a remarkable time gap between the 16-th and 17-th temple consecration.   
 After 1923, a smooth growth could resume, as seen in Figs.1-2, with a lower but nevertheless appreciable rate. A remark should be made here: observe that there is not much effect  $ca.$ 1929, i.e. at the most famous financial crash time before the "Great Depression". 

 I wish to emphasize  what seems to be a (financial) paradox in such a social force concept:  the sect adepts were likely under much hardship due to the 1-st World War. Nevertheless, in  need of some intra-community social support,   the adepts donated quite an amount of money to build  temples. Very interestingly, it has been found in a modern context that  a similar situation, i.e. giving more at bad times,  has just been occurring in Ireland \cite{TimeMag}. This is a noticeable example of some sort of true altruism, in contrast to results from fund-raising campaigns  \cite{QJE127.12.1altruism}.

 \begin{figure}
\centering
\includegraphics[height=14 cm,width=16cm]{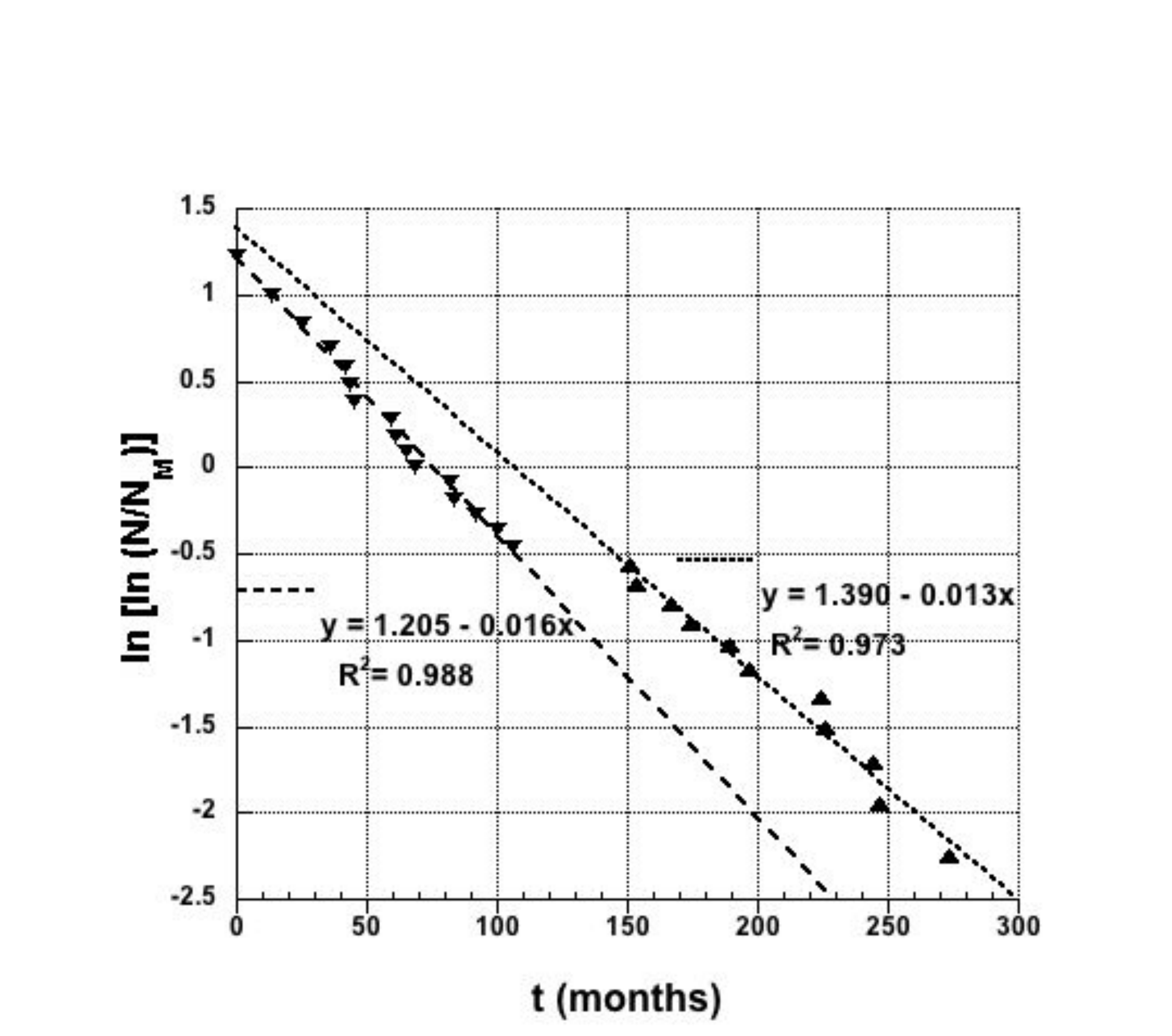}
\caption{Gompertz plot : Log(Log) variation of the number of temples ($N$) raised  in Belgium as a function of the number of months, starting from the raise of the first temple in Jemeppe-sur-Meuse, indicating the effect of a "social force" influencing a variation in growth rates, starting  at $t \sim$ 150. }
\label{fig:Plot59templeGforcellt}
\end{figure}

\section{Conclusions}\label{Conclusions}

As an initial remark for a conclusion, let it be noted that both Verhulst and Gompertz  first order differential equations for the evolution of a population  are peculiar approximations of a more general one, i.e.
\begin{equation}\label{MathBioeqTW}
\frac{dN}{dt}= rN^{a}  \biggl[1- \big(\frac{N}{M}\big)^{b} \biggr ]^{c}
\end{equation}
where $a$, $b$, $ c$  can be requested to be positive real numbers; $M$ is again the carrying capacity.  Typical features on the relative and maximum  growth rates, asymptotic values, inflection point characteristics are easily derived  \cite{MaBio179.02.21}. 
 Introducing $x \equiv (N/M)^{b}$, one can rewrite Eq.(\ref{MathBioeqTW}) as
\begin{equation}\label{MathBioeqTWx}
\frac{dx}{dt}= \frac{b\; r\;}{ M^{1-a} }\; \; x^{1+(\frac{a-1}{b})} \; [1-x]^{c}.
\end{equation}
Obviously, Verhulst equation is Eq.(\ref{MathBioeqTW}) or Eq.(\ref{MathBioeqTWx}) when    $a$=$b$=$ c$=1. Gompertz growth equation is obtained for   $a$=$c$=1, and $b\rightarrow$0.

To introduce the Montroll's form, before applying a pulse force, one rewrites the above equation as
\begin{equation}\label{MathBioeqTWxM}
\frac{d\;ln(x)}{dt}= \frac{b\; r\;}{ M^{1-a} }\;\;  x^{(\frac{a-1}{b})}\;  [1-x]^{c}.
\end{equation}
Thus, as long as the "social force" term has a structure similar to that of the right hand side, i.e. a power of $x$ $or$ of $1-x$, the integration can be easily performed., for example,  in terms of {\it  incomplete  Beta functions}. 
This idea can be carried over  within the Gompertz framework. It is "sufficient" to replace $(1-x)$ by $\sim -ln(k/y)$; see Eq.(\ref {Gompertz1dif}).


 
   In  summary,  this study   differs from   that of Montroll who claimed that the evolution of some product results from a competition effect  between two processes.   He imposes a change in the $x$  (or $X$)term, i.e. the already acquired content. In fact,  Montroll's adaptation of Verhulst equation is more like an economic field competition description  rather than a "social force".      In the present considerations, the new term, called by analogy with Montroll's paper, a  "social force" is attributed to a change in the  limited need/capacity, in Verhulst or Gompertz equation,  i.e. with emphasis on the $(1-X)$ term.
The  change in evolution is rather due to an external field, through the coefficient in front of $(1-X)$.

  In the studied example, the gap between the first and second regime has  no doubt some extrinsic effect, i.e.  an expected parliamentary law inducing a tax free system for the community,  thereby hindering the temple constructions for a while. 
  Thus, it can be pretended that  one can be  presenting two different views of a "social force":  one resulting from  a competitive interaction and another from an   external field.  

Furthermore, note that  the formal introduction of an extra $(1-X)$ term  contribution rather than an $X$ term based "pulse", as in Montroll's paper, not only allows for greater flexibility in the data fit, but also in the analysis and  interpretation of the resulting graphs. Indeed, the $(1-X)$  term  adaptation allows for smoothness in the evolution, over some time interval, rather than requesting a jump in the evolution, with two constant  and identical rates. 
Furthermore, it seems, "sociologically interesting", that the $X$ term, like in Malthus approach bears upon what exists already, but the   $(1-X)$ term, like in the Verhulst approach, bears upon what is missing.

In conclusion, based on  a relatively simple analysis of reliable data indicating adhesion to a growing sect, i.e. the construction of temples, a social force effect, in the physics sense introduced by Montroll, can be observed.  However, a different point of view can be taken. Moreover,   comparing Verhulst and Gompertz law, it  has been found that although both growth laws are   based on different empirical considerations, the laws are found to be quite complementary. 
 From a sociological point of view of such complex systems, the "model" indicates that such communities are markedly influenced by external considerations ("external  fields/forces"), beside intrinsic "religious" goals.

 \bigskip

{\bf Acknowledgements} 
 \bigskip


 Comments on preliminary  considerations  by C. Berman,  J. Hayward, A. P\c{e}kalski, and F. Schweitzer, and later on by anonymous reviewers,  have surely improved this paper.


\begin{thebibliography}{99}
 
\bibitem{PNAS75.78.4633-7-Montroll-socialforces}  Montroll, E.W.,
 Social dynamics and the  quantifying of social forces,
   {\it Proc. Nat.  Acad.  Sci.  USA} {\bf 75} (1978)  4633--4637.
  
\bibitem{Malthus798} Malthus,  T. R., An Essay  on the Principle of Population
as  It Affects the Future Improvement of Society (J. Johnson, London, 1798).

  \bibitem{Verhulst845}
 Verhulst, P.F.,
Recherches math\'ematiques sur la loi d'accroissement de la population, 
 {\it   Nouveaux M\'emoires de l'Acad\'emie Royale des Sciences et Belles-Lettres  de Bruxelles}   {\bf 18} (1845) 1-38. 
  
\bibitem{Verhulst847}
  Verhulst, P.F.,  {\it Deuxi\`eme m\' emoire sur la loi d'accroissement de la
population}. {\it Nouveaux m\' emoires de l'Acad\'emie Royale des Sciences et Belles
Lettres de Bruxelles}, {\bf 20} (1847)  1-32.

   \bibitem{Gompertz825} Gompertz,  R., On the Nature of the Function Expressive of the Law of Human Mortality, and on a New Mode of Determining the Value of Life Contingencies, {\it Philos.  Trans. R.  Soc. London}, {\bf 115} (1825) 513-585.
  

 




 
 

 


   
     
\bibitem{MontrollBadger74}     Montroll, E.W. and  Badger, W. W., {\it  Introduction  to Quantitative Aspects of Social Phenomena}   (Gordon and Breach, New York, 1974). 

\bibitem{Helbing1}   Helbing, D.,   A mathematical model for the behavior of individuals in a social field,   {\it  J.  Math. Sociol.} {\bf 19} (1994) 189-219.

\bibitem{Helbingbook}    Helbing,  D.,  {\it  Quantitative sociodynamics: stochastic methods and models of social interaction processes} (Springer, Berlin, 2011), pp. 225-245.
 

  

 
  
\bibitem{religion1} Ausloos, M. and  Petroni, F., Statistical dynamics of religions and  adherents,   {\it Europhys. Lett.} {\bf 77} (2007) 38002.

\bibitem{mendes}  Picoli, S., Jr., and   Mendes, R. S. ,  Universal features in the growth dynamics of religious activities, {\it      Phys. Rev. E } {\bf  77} (2008) 036105.

\bibitem{religion2} Ausloos, M.  and  Petroni, F.,   Statistical dynamics of religions, 
{\it Physica A} {\bf   388} (2009)  4438--4444. 
%
 
\bibitem{religion568} Ausloos, M.,  On religion and language evolutions seen through  mathematical and agent based models,  in     {\it Proceedings of the First Interdisciplinary CHESS  Interactions Conference}, eds. Rangacharyulu, C. and Haven, E.,  (World Scientific,  Singapore, 2010)  pp. 157-182.

%
\bibitem{hayw99} Hayward, J.,  Mathematical modeling of church growth,
   {\it  J. Math. Soc.} {\bf 23} (1999) 255--292. 
%
\bibitem{hayw05}  Hayward, J., A general model of church growth and decline, 
 {\it  J. Math. Soc.} {\bf 29} (2005) 177--207.
   

\bibitem{MAarXiv1201.4841}  Ausloos, M., Econophysics of a religious cult: the Antoinists in Belgium [1920-2000],  {\it   Physica A }   {\bf 391}  (2012) 3190-3197 .


\bibitem{MaBio179.02.21}  Tsoularis, A.  and  Wallace, J., Analysis of logistic growth models,
{\it    Math. Biosci.}, {\bf 179} (2002) 21-55.

          
\bibitem{Debouxhtaybook}   Debouxhtay, P., {\it Antoine le Gu\' erisseur et l'Antoinisme} (Gothier, Li\`ege, 1934).

  \bibitem{Dericquebourg}   Dericquebourg, R., {\it Les Antoinistes} (Brepols, Maredsous, 1993).
            
 
\bibitem{chi2table} {\it http://home.comcast.net/$\sim$sharov/PopEcol/tables/chisq.html}; 

 {\it http://www.itl.nist.gov/div898/handbook/eda/section3/eda3674.htm}  
 
\bibitem{2280354fits}   Brennan, J.F., Evaluation of Parameters in the Gompertz and Makeham Equations,  	 {\it   J. Am. Stat. Ass. } {\bf  44} (1949) 116-121. 

 \bibitem{2280430fits Farebrother}    Sherman, J. and Morrison,  W.  J., Simplified Procedures for Fitting a Gompertz Curve and a Modified Exponential Curve,    {\it   J. Am. Stat. Ass. } {\bf  45} (1950) 87-96.

 \bibitem{Norton76NatureGompertzgrow}   Norton, L.,  Simon, R.,  Brereton, H.D. and  Bogden, A.E.,  Predicting the course of Gompertzian growth,   {\it   Nature } {\bf   264} (1976) 542-545.
 
  \bibitem{2347021comparingVG}   Vieira, S. and  Hoffmann, R.,    Comparison of the Logistic and the Gompertz Growth Functions Considering Additive and Multiplicative Error Terms,  {\it    J.  Royal Stat.  Soc.  C } {\bf  26} (1977) 143-148.
 
    \bibitem{3315118fits}   Ramasubban, T. A.,  A Comparison of Alternative Rate Estimators in Some Frequently Used Growth Models,   {\it  Can. J. Stat./   Rev. Can. Stat. } {\bf 7} (1979) 185-192.
    
  \bibitem{2583955fits}   Franses, P.H., Fitting a Gompertz Curve,   {\it   J. Oper. Res.  Soc. } {\bf  45}  (1994) 109-113.
 
\bibitem{PhA368.06.225Gompertzrosu}   Ibarra-Junquera, V.,   Monsivais, M.P.,   Ros\`u, H.C. and  L\' opez-Sandoval, R.,   A robust estimation of the exponent function in the Gompertz law,    {\it   Physica A } {\bf 368} (2006) 225-231.
  
 
  \bibitem{footnote} Montroll hoped that, as the collection
of  $k$ and $\alpha$ values grows, one  could develop some intuition on the
evolutionary rate constants associated with various classes of
processes and the influence of various types of accelerating and
deterring forces. This is still to be done.  


\bibitem{TimeMag} Gibson, M.,  A Tale of Two Donors:
Cash-Strapped Ireland Outshines
Germany in Humanitarian Aid, Time Mag.,  Friday Mar. 23, 2012;
$http://www.time.com/time/world/article/0,8599,2109903,00.html$.

\bibitem{QJE127.12.1altruism} DellaVigna, S., List, J.A., Malmendier, U., Testing for altruism and social pressure in charitable giving,    {\it  Quarter. J. Econ.} {\bf 127} (2012)  1-56.

\end{thebibliography}
\end{document}